# Superconductivity in the A15 Structure

G. R. Stewart

Department of Physics, University of Florida, Gainesville, FL  32611

**Abstract:  The cubic A15 structure metals, with over 60 distinct member compounds,  held the crown of highest $T_c$ superconductor starting in 1954 with the discovery of $T_c$=18 K in $Nb_3Sn$.  $T_c$ increased over the next 20 years until the discovery in 1973 of $T_c$=22.3 K (optimized to ≈23 K a year later) in sputtered films of $Nb_3Ge$.  Attempts were made to produce - via explosive compression - higher (theorized to be 31-35 K) transition temperatures in not-stable-at-ambient-conditions A15 $Nb_3Si$.  However, the effort to continue the march to higher $T_c$'s in A15 $Nb_3Si$ only resulted in a defect-suppressed $T_c$ of 19 K by 1981.  Focus in superconductivity research partially shifted with the advent of heavy Fermion superconductors ($CeCu_2Si_2$, $UBe_{13}$, and $UPt_3$ discovered in 1979, 1983 and 1984 respectively) and further shifted away from A15's with the discovery of the perovskite structure cuprate superconductors in 1986 with $T_c$=35 K.  However, the A15 superconductors - and specifically doped $Nb_3Sn$ – are still the material of choice today for most applications where high critical currents (e. g. magnets with dc persistent fields up to 21 T) are required. Thus, this article discusses superconductivity, and the important physical properties and theories for the understanding thereof, in the A15's which held the record $T_c$ for the longest time (32 years) of any known class of superconductor since the discovery of $T_c$=4.2 K in Hg in 1911.  The discovery in 2008 of $T_c$=38 K at 7 kbar in A15 $Cs_3C_{60}$ (properly a member of the fullerene superconductor class), which is an insulator at 1 atm pressure and otherwise also atypical of the A15 class of superconductors, will be briefly discussed.**

I.      **Introduction:**

Until the discovery [1] by Hardy and Hulm of 17.1 K superconductivity in cubic A15 structure $V_3Si$ in March 1954, the cubic NaCl structure class of materials had had no competition for the record highest $T_c$. The upwards climb of $T_c$ in the NaCl structure materials began with $T_c$=10.3 K in [2] 1933 for NbC, followed by 15.25 K in [3] NbN in 1942 [4] (15.98 K in 1952 [5]).  Matthias reported [6] (essentially at the same time as the discovery of superconductivity in $V_3Si$) $T_c$=17.8 K in November 1953 for $NbC_{0.3}N_{0.7}$, but the record $T_c$ passed to the A15's in September 1954 (for [7] $Nb_3Sn$, $T_c$=18.05 ± 0.1 K) and stayed with the A15's until 1986.  There were other 'high' $T_c$ materials discovered during this period (e. g. bcc $Pu_2C_3$ structure $Y_{0.7}Th_{0.3}C_{1.5}$, $T_c$=17 K in [8] 1969), but A15's were by far the much larger class and the main focus in the search for higher $T_c$ during this period. After the discovery of what was at the time 'high temperature' superconductivity in $V_3Si$ and $Nb_3Sn$ only six months apart, the search for other examples in the A15's with higher $T_c$ did not progress for more than a decade.  Then, $T_c$ was found to be 20.0 K in $Nb_3Al_{0.8}Ge_{0.2}$ in [9] 1967, 18.8 K in $Nb_3Al$ in [10] 1969 (previously 18 K [11] 1959), 20.3 K in $Nb_3Ga$ in [12] 1971 and finally 22.3 K in $Nb_3Ge$ in [13] 1973, optimized to 22.9 K in [14] 1974 (23.2 K in ref. 15).

This article is intended to give an overview of the A15 class of superconductors, which (despite being bypassed in the quest for higher $T_c$ by the cuprates in 1986, by $MgB_2$ in 2001, and by the iron based superconductors in 2008) remain the leader in applications (e. g. medical imaging) requiring magnets with fields larger than 10 T.  Considered to be conventional, BCS superconductors, the study of the A15's led to important insights as to the causes of electron-phonon mediated superconductivity and also progress in materials preparation and characterization which has been useful in studying and applying the succeeding classes of superconductors.

For ease of navigation for the reader, the discussion on A15's in section II below is divided into five sections: 1. materials preparation and properties/structure/applications; 2. theoretical understanding of why $T_c$ is so high; 3. important properties: resistivity, susceptibility, specific heat, upper critical field, other; 4. attempts to go past $T_c$=23 K in $Nb_3Ge$: A15 $Nb_3Si$; 5. comparison of the conventional A15 superconductors with other classes of superconductors and summary.

## II. Discussion of A15's as a class of superconductor

### 1. A15's from a materials perspective

The cubic A15 structure, pictured in Figure 1, is also called $\beta$-W, since the first observation of the structure in 1931 was in an allotrope of tungsten. The prototypical A15 compound is the non-superconducting $Cr_3Si$. Although there are often variations of stoichiometry, the ideal formula unit is $A_3B$, where A is a transition metal like V, Nb, or Mo and B is from the right side of the periodic table, including elements like Al, Si, Ge and Sn. Some examples of A15's have stoichiometries far from the canonical A$_3$B, e. g. Mo$_{0.4}$Tc$_{0.6}$ (T$_c$=13.4 K [18]) and V$_{0.29}$Re$_{0.71}$ (T$_c$=8.4 K [19]) with B atoms on the A-sites, and Nb$_3$(Nb$_{0.92}$Ge$_{0.08}$) or 'Nb$_3$Nb' stabilized in the A15 structure by a few percent Ge, T$_c$=5.2 K [20], with A atoms on the B-sites. In V$_3$Ga, the A15 structure phase extends [21] from 18 to 32 % Ga, with however the highest T$_c$ (14.5-15 K) at the stoichiometric 25% composition and a sharp fall off in T$_c$ (approximately a factor of two for a change in Ga composition of ±5%) away from this 3:1 stoichiometry [21]. The history of the efforts to increase T$_c$ in the A15's after superconductivity in V$_3$Si and Nb$_3$Sn was discovered in 1954 is essentially a history of struggling to achieve the proper 3:1 stoichiometry in compounds where the A15 structure was not stable there, i. e. in $Nb_3Ga$, $Nb_3Ge$, and $Nb_3Si$. Matthias et al., in their early work on $Nb_3Ge$, stated [22] "It is always the stoichiometric [A15] compound which has the maximum transition temperature." (As will be discussed in section II.2 (theoretical understanding) lattice disorder – including mixing atoms on a particular sublattice - strongly affects the electronic density of states and thereby $T_c$.)

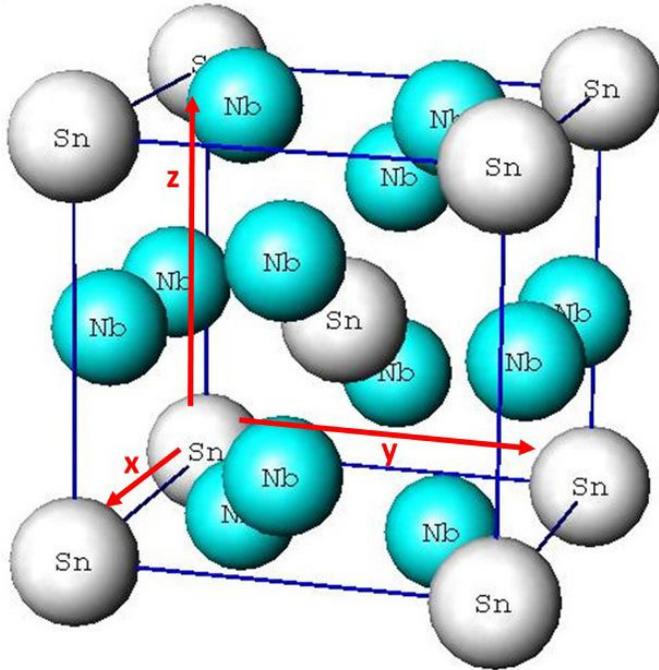

Fig. 1: (color online) Diagram [16] of A15 Nb$_3$Sn, which has a cube edge (lattice parameter '$a_0$') of length 5.29 Å. The B atoms form a body centered cube, and the A atoms form one dimensional chains in the three orthogonal directions, with an interatomic spacing along the chains of ½ of the lattice parameter. For A15 Nb$_3$Sn, this gives a Nb-Nb spacing much closer (7.5 %) than in, e. g., pure Nb which has the highest elemental T$_c$ at 9.2 K. This rather unique structure has an important influence on the physical properties (including electronic density of states at the Fermi energy, N(0), and the phonon spectrum), as will be discussed in the theoretical understanding section. Some samples of Nb$_3$Sn and V$_3$Si exhibit [17] a martensitic phase transformation from cubic → tetragonal upon cooling, discussed below in section II.2.

The two highest known $T_c$ metallic A15's, $Nb_3Ga$ and $Nb_3Ge$, will now be discussed to illustrate the difficulty achieving 3:1 stoichiometry and the maximum $T_c$, with 13 years being required to attain optimal $T_c$ in $Nb_3Ga$ and 17 years required in the case of $Nb_3Ge$, which is unstable in bulk form and was finally stabilized at 3:1 in the A15 structure in thin film form by sputtering.

Matthias and co-workers reported [23] $T_c$=14.5 K for nominal $Nb_3Ga$, $a_0$=5.171 Å, in 1958, with no special effort given to determine the actual stoichiometry. Webb et al. [12] in 1971 succeeded (after great effort) is preparing essentially stoichiometric $Nb_3Ga$, $T_c$=20.3 K, (the first reported binary compound with $T_c$>20 K) with the lowest lattice parameter ever reported for this compound, 5.165 Å. They found a monotonic rise of $T_c$ in $Nb_3Ga$ with decreasing lattice constant, $a_0$, where the smaller $a_0$ is simply a metric for the approach to the perfect 3:1 stoichiometry. This point (that the $T_c$ increase is due to the approach to unbroken chains of A-atoms and is not caused by the decrease in interatomic spacing) is made clear by the measurement [12] of a *depression* of the superconducting $T_c$ in the $T_c$=14.5 K $Nb_3Ga$ material under pressure. See ref. 24 for an overview on work on $Nb_3Ga$, where $T_c$ was eventually increased to 20.7 K.

The success of Gavaler to achieve stoichiometric $Nb_3Ge$ and $T_c$'s approaching 23 K was the culmination of a community wide effort based on well-established trends of $T_c$ values in the A15's with lattice constants. It was known that $T_c$ was inversely proportional to lattice parameter in a given A15 family like $Nb_3B$ where B is isoelectronic, i. e. in the same column in the periodic table. For example, B=In, $a_0$=5.303 Å, $T_c$=9.2 K; B=Al, $a_0$=5.182 Å, $T_c$=18.8 K; B=Ga, $a_0$=5.165 Å, $T_c$=20.7 K. $T_c$ is also $\propto 1/a_0$ within a specific compound like $Nb_3Ga$ or $V_3Ga$ where $T_c$ has been studied as a function of lattice parameter. Thus, since the ionic radius of Ge (1.37 Å) is much smaller than that of Sn (1.62 Å), the expectation was that $T_c$ for $Nb_3Ge$ would be significantly larger than the 18.05 K $T_c$ for $Nb_3Sn$. (The search for even higher $T_c$ in A15 $Nb_3Si$, where the ionic radius for Si is 1.32 Å, is discussed below in section II.4.)

The efforts to achieve higher $T_c$ in $Nb_3Ge$ started rather humbly. Carpenter and Searcy [25] reported $a_0$=5.168 ± 0.002 Å in 1956 for '$Nb_3Ge$', and $T_c$ was reported [26] to be 6.90 K in 1963. From there, Matthias et al. [22] in 1965, motivated by the observation by Geller [27] that the proper lattice parameter for stoichiometric $Nb_3Ge$ should be 5.12 Å, prepared '$Nb_3Ge$' with a broad superconducting transition (starting at $T_c^{onset}$=17 K and extending down to 6 K) using a rapid quench technique. The lattice parameter achieved in the somewhat disordered alloys, with almost half of the Ge atoms on the 1 dimensional chain Nb sites, was $a_0$=5.149 ± 0.005 Å. By 'splatting' a molten mixture of Nb and Ge with Ge in excess of 25%, the idea was to increase the inter-solubility of the two elements. However, the thermal quenching led [22] to unavoidable site disorder, which was known to lower $T_c$ in the A15's.

Eight years later, in 1973, Gavaler [13] at Westinghouse R & D succeeded in sputtering thin (1μm) films of metastable $Nb_3Ge$ on a heated substrate with $T_c^{onset}$ =22.3 K, transition width only 1.5 K, and $a_0$=5.15 ± 0.01 Å. In succeeding work [14], Gavaler, Janocko and Jones found reproducible $T_c^{onset}$ values of 22.4 K (with transition widths of 0.7 K and $a_0$=5.143 ± 0.003 Å) and some samples with $T_c$ as high as 22.9 K. (Testardi et al. [15], were able to quickly duplicate the method of Gavaler [13] and reported $T_c$ as high as 23.2 ± 0.2 K.) The slope of the upper critical field values at $T_c$, $dH_{c2}/dT|_{Tc}$, for these early films from Gavaler was measured to be approximately = 2.4 T/K and the extrapolated upper critical field, $H_{c2}(0)$, was [28] ≈ 37 T. Much work was done (e. g. see ref. 29) in the succeeding decade to prepare application-capable $Nb_3Ge$ films by sputtering and <u>C</u>hemical <u>V</u>apor <u>D</u>eposition (CVD). (See also discussion of the upper critical field measurements on A15's in section 3 below.)

However, modern high field magnets are produced using several optimizations of the more stable $Nb_sSn$. Alloying of Ta (4%) or Ti (2%) with the Nb increases [30] $H_{c2}(0)$ in $Nb_3Sn$ by approximately 3.5 T and $T_c$ by ≈ 0.3 K. Wire is produced by varying processes, including a bronze process in which Nb rods are placed in a pattern in a Cu-Sn bronze matrix with pure Cu surrounding for thermal stability (see Fig. 2). The entirety of this is then drawn down to the desired wire diameter. This composite wire is then wound on the magnet solenoid, and only then reacted in place to form the $Nb_3Sn$ on the surface of the Nb filaments by diffusion due to the brittleness of the $Nb_3Sn$ A15 conductor. The superconducting wire is placed under tensile stress upon cooling to liquid helium temperatures for operation as a magnet solenoid for a further increase in upper critical field. Such wire has $H_{c2}(T=0)$ values of 29.5 T and a $T_c$ of 17.8 K in practical, long length conductors [31]. Actual magnets in production (e. g. at Oxford Instruments) reach 22.3 T at 2.2 K (superfluid helium temperature). Such commercial Nb$_3$Sn magnets will be surpassed at some point by cuprate superconductor magnets operable above 30 T currently in prototype development stage.

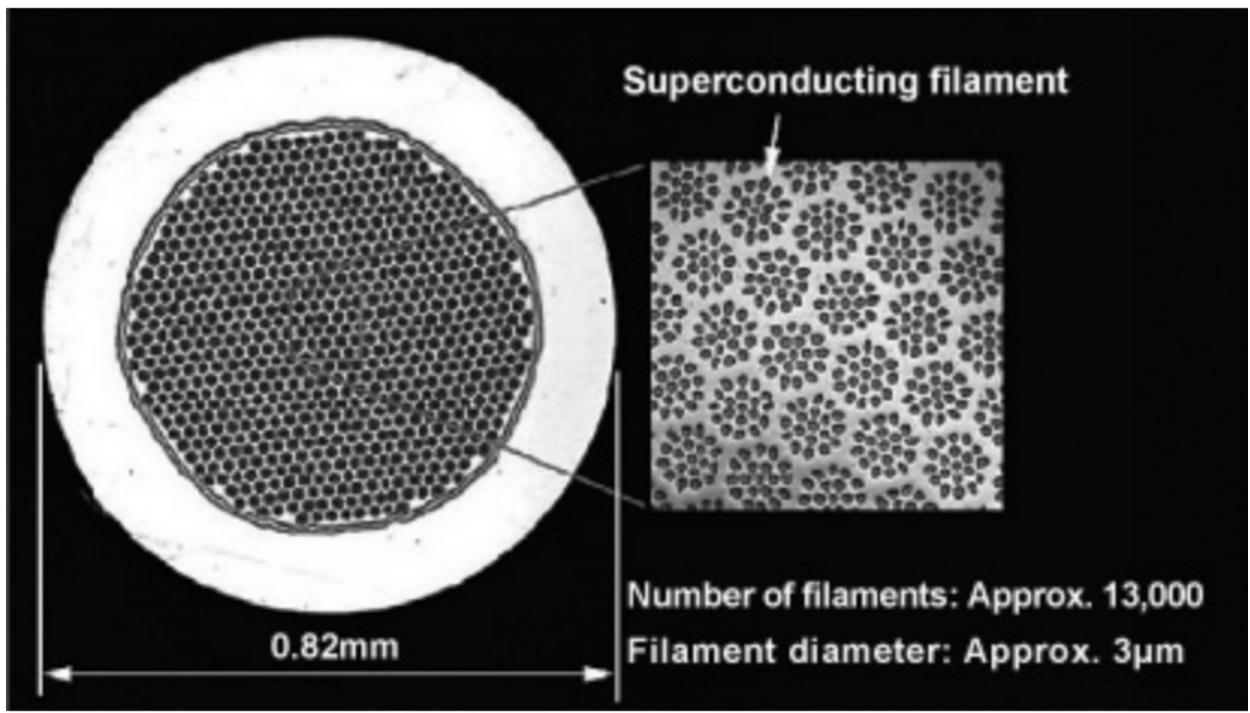

Fig. 2: Multifilamentary Nb$_3$Sn wire produced by Japan Superconductor Technology, Inc. The expanded view on the right shows ≈25 bundles of 19 filaments each.

2. <u>Theoretical Understanding of Superconductivity in the A15's</u>

This is a rather broad topic. A15 superconductors are considered to be describable by the BCS theory, i. e. the pairing of the superconducting electrons is via electron-phonon coupling. Thus, the phonon spectrum, the electronic density of states at the Fermi energy, and the coupling between the electrons and the phonons are discussed as determining T$_c$.

As stated above, the 1-dimensional chains of atoms of transition metals like Nb or V, with reduced inter-atomic spacing vs the pure element, are characteristic of the A15 structure and influence/increase the electronic density of states, N(0), at the Fermi energy. In weak coupling BCS theory, $T_c \propto <\omega>\exp(-1/N(0)V)$ where $T_c$ is proportional to an average phonon frequency, $<\omega>$, the density of states at the Fermi energy, and an electron-phonon coupling parameter V (often also characterized by the parameter $\lambda$). Due to the long reign of the A15's as the highest known $T_c$ materials, the belief that high $N(0) \Rightarrow$ high $T_c$ became quite ingrained. It was this assumption that Bednorz and Mueller eschewed to find superconductivity in the cuprates.

McMillan proposed [32] an improved, partially phenomenological $T_c$ equation for strong coupled superconductors (which the higher $T_c$ A15's certainly are), $T_c=(\Theta_D/1.45)\exp(-[1.04(1+\lambda)]/[\lambda-\mu^*(1+0.62\lambda)])$, with $\lambda$ the electron-phonon coupling parameter (discussed further below), $\mu^*$ is the Coulomb coupling constant, and $\Theta_D$ is the Debye temperature. For weak coupling, $\lambda<<1$, this formula goes over into the weak coupled BCS one with $\lambda-\mu^*$ replacing N(0)V.

Early pioneering work [33] on calculating the electronic structure of the A15's was carried out by Mattheiss in the mid 1960's using augmented plane wave techniques. Calculations of N(0) progressed markedly in the 1970's, with improved computer codes and methods. As an example, Pickett, Ho, and Cohen [34] used a self-consistent pseudopotential method in 1979 to calculate the band structure and N(0) for the A15 compounds $Nb_3Ge$, $Nb_3Al$, and theoretical $Nb_3Nb$ (one year before this compound was experimentally realized [20] with a slight amount of Ge to stabilize the

A15 structure.) These calculations put the Fermi energy just in a range where very flat bands (energy, E, approximately constant with wave vector, k) occur, giving high N(0) ($\propto 1/(dE/dk)$). For $Nb_3Ge$, the Fermi energy was found to lie at the center of a peak (width $\approx$ 0.06 eV) in the density of states vs energy; N(0)=7.6 states/(eV-spin-unit cell). For $Nb_3Al$, the Fermi energy was found to lie on the edge of a 'huge peak' in the density of states, width $\approx$ 0.15 eV; N(0)=7.8 states/(eV-spin-unit cell). Very recent work [35] (N(0) $\approx$ 7.2 states/(eV-spin-unit cell for both $Nb_3Ge$ and $Nb_3Al$) using full potential linearized augmented plane wave calculations essentially agrees with this 35 year old result– a thorough vote of confidence for the earlier result. For theoretical $Nb_3Nb$, the Fermi energy was found in ref. 34 to lie between two large peaks; N(0)=4.1 states/(eV-spin-unit cell).

Since the $T_c$ values for A15 $Nb_3Ge$ and $Nb_3Al$ are similar in magnitude (22.9 and 18.9 K respectively), and that for A15 '$Nb_3Nb$' is much smaller ($\approx$ 5.2 K), the results of the band structure calculations for N(0) seem at least qualitatively consistent with the premise that higher N(0) brings higher $T_c$ values. As a comparison, the band structure calculation [36] for bcc elemental Nb, $T_c$=9.2 K, results in an N(0) that is $\approx$ 40% larger [37] than for A15 $Nb_3Nb$ (consistent with the larger $T_c$), but however not that dissimilar to the values for A15 compounds with $T_c$ values above 17 K (Table 1 below). Thus $T_c$ scaling with N(0) is at best a qualitative metric.

Therefore, clearly the phonon spectrum and the electron-phonon coupling must also be considered. $T_c$ – despite phenomenological thinking encouraged by the closeness/increased orbital overlap along the 1 dimensional chains of transition metal atoms in A15's - is not just proportional to N(0). Ho, Cohen, and Pickett pointed out [40] that large electron-phonon coupling (beneficial in the BCS theory for higher $T_c$) has the effect in the A15 compounds of smearing out sharp features in the electronic density of states, thus having an unexpected negative influence on $T_c$. As an example, they estimate that the calculated electron-phonon coupling parameter $\lambda$ in $Nb_3Ge$ will so smear out the calculated narrow (0.06 eV) peak at the Fermi energy that the low temperature effective N(0) will be reduced 20-30%, having a 'drastic effect' on $T_c$. This makes even clearer the importance of considering all factors together in trying to understand the 'high' transition temperatures in the A15's.

Certainly there is ample precedent in studying superconductivity in the A15's for considering the phonons to be important. As mentioned above, two of the higher $T_c$ A15 compounds, $Nb_3Sn$ and $V_3Si$, were known to undergo extreme phonon softening in certain modes (the elastic modulus $c_{11} - c_{12}$ goes to zero) leading to a martensitic (volume conserving) transition (see Fig. 3) at temperatures $T_M$ rather close to, but above, $T_c$ – 44.5 and 20.5 K respectively. A fair amount of work was devoted trying to find a link between this phonon softening and superconductivity, see the review [41] by Testardi. Although there were a number of reports of *indications* of cubic → tetragonal transformations in additional A15 compounds (e. g. see the discussion in ref. 34) in the end only $Nb_3Sn$ and $V_3Si$ showed convincing evidence (low temperature x-ray or neutron scattering data or bulk anomalies in the low temperature specific heat) – and that only in a subset of the samples. In non-transforming single crystal $V_3Si$ the elastic modulus $c_{11} - c_{12}$ still falls by 85% between room temperature and $T_c$, where further softening is arrested by the superconducting transition. [41] In measurements on polycrystalline $Nb_3Al$ and $V_3Ga$, although there is no transformation, there is still a lattice softening observed via a decrease in the sound velocity (by 2 and 4 % respectively) between 300 and 20 K. [45]

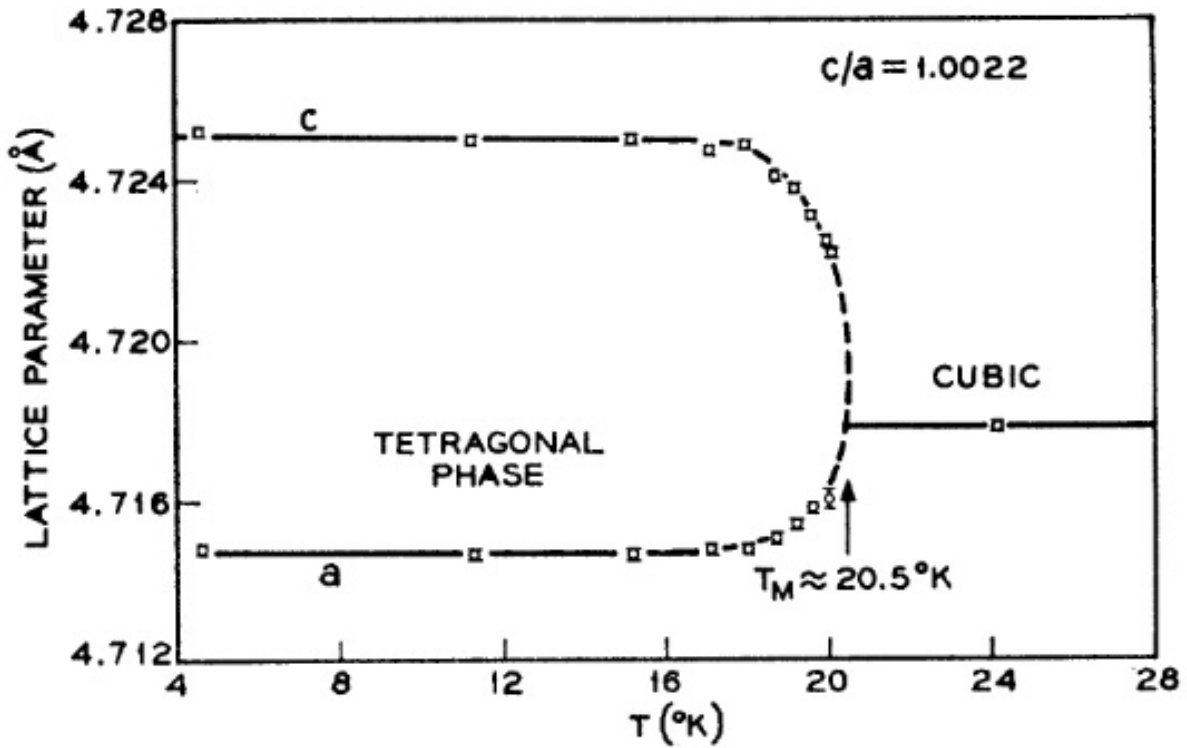

**Fig. 3:** Depiction of the second order martensitic transition [17] in $V_3Si$, where due to acoustic phonon mode softening the cubic unit cell transforms to tetragonal upon cooling. The volume of the cell remains constant. In $Nb_3Sn$ c/a becomes [42] less than 1 below the martensitic transformation, which – in a further contrast to $V_3Si$ – is [43] first order. Whether improved measurements would determine the cubic-tetragonal transition in $V_3Si$ to be also of first order remains an open question. [44]

Thus, the idea that the higher $T_c$ A15's were characterized by mode softening and that this was important for the transition temperature had its proponents. Obviously, the full collapse of the lattice stiffness in a particular direction was not the central issue, since non-transforming and transforming single crystals of $V_3Si$ and $Nb_3Sn$ have essentially identical $T_c$'s. Instead, the tendency towards structural instability and the associated lattice softening (decrease in average phonon frequency $<\omega^2>$) was thought (for an early review see ref. 41) to play a role through an enhanced electron-phonon coupling $\lambda$.

In a review [46] of electron phonon coupling effects by Pintschovius (see also ref. 47), the point is made that $Nb_3Sn$ ($T_c$=18 K) shows distinct phonon anomalies and phonon softening, while the low $T_c$ $Nb_3Sb$ does not. Phonons with anomalously low frequencies often show [46] an anomalous temperature dependence, softening upon cooling rather than exhibiting the usual slight hardening related to anharmonicity. Such a behavior is observed [46] for the $Nb_3Sn$ longitudinal acoustic branch. Theory predicts [46] that this anomalous softening upon cooling should be accompanied by relatively large neutron scattering linewidths, which are directly related to the electron–phonon coupling constant $\lambda$. Unfortunately in A15 $Nb_3Sn$, such line width broadening is at the edge of experimental resolution. (This is not [46] the case in the 39 K, electron-phonon coupling superconductor $MgB_2$.)

Thus, a picture of the precise relative impacts of N(0), $\lambda$ (discussed more in section 3, specific heat, below), and phonon softening on $T_c$ in the A15's is a subject for exact calculation, with all three playing an entertwined role.

3. **Important physical properties:** Resistivity ($\rho$), Magnetic Susceptibility ($\chi$), Specific Heat (C), Upper Critical Field ($H_{c2}$), Other

$\rho$: The unusual phonon properties of the A15 superconducting class also have their influence on the physical properties. Just as the high $T_c$ cuprate materials show unusual normal state resistivity ($\rho \propto T$ up to 1100 K in $La_{1.825}Sr_{0.175}CuO_4$, see ref. 48), so – in a different way – do the A15 superconductors. $\rho \propto T^2$ from $T_c$ up to ~50 K [49] for $Nb_3Sn$, $Nb_3Al$, and $Nb_3Ge$ while for temperatures up to 800 K $\rho$ for the higher $T_c$ A15's shows [50] negative curvature. For polycrystalline $Nb_3Sn$, ref. [51] reported $\rho = \rho_0 + \rho_1 T + \rho_2 \exp(-T_0/T)$, with $\rho_0 \approx 10$ $\mu\Omega$-cm (i. e. a relatively good metal) and $T_0 = 85$ K. The explanations put forward [50] for this anomalous behavior involves either the sharp structure in N(0) or the anharmonic hardening of the phonon modes as temperature increases. (For the cuprates, calculations [52] showed that $\rho \propto T$ is expected from a proper consideration of electron-phonon scattering and the measured phonon spectrum.) For $V_3Si$ – the only other A15 preparable with essentially perfect sublattice order like $Nb_3Sn$ – a $\rho_0$ value of 0.9 $\mu\Omega$-cm and a residual resistivity ratio (RRR), $\rho(300 K)/\rho(T \rightarrow 0 K)$), of 84 in optimized single crystals has been reported [53].

$\chi$: Measurements of the magnetic susceptibility $\chi$ and the Knight Shift, K, in nine $V_3X$ A15's showed [54] an interesting correlation between the size of the temperature dependence (for example, $\chi$ in $V_3Ga$, $T_c \approx 15$ K, increases by 50% upon cooling from 300 to 10 K while K decreases by 20%) and $T_c$. For low $T_c$ A15 $V_3X$, e. g. $V_3Au$ ($T_c = 0.7$ K), there is essentially no change in $\chi$ and K. Clogston and Jaccarino [55] proposed early on (in 1961) a model to account for this anomalous temperature variation in $\chi$ and K in $V_3X$ compounds assuming a sharp peak in the density of states close to $\varepsilon_{Fermi}$. Labbé and Friedel in 1966 [56] derived peaks in $N(\varepsilon)$ based on the orthogonal linear chains of transition metal d-electron atoms (Fig. 1) which could explain [54] the behavior of $\chi$ and K. As discussed already in section 2 just above, more modern computerized band structure calculations also find sharp structure in $N(\varepsilon)$.

There is also large temperature dependence in $\chi$ (increase between 300 and $T_M$ of 30% [57]) and in K [58] in $Nb_3Sn$ like that seen [54] in the high $T_c$ $V_3X$, which is consistent with calculations of sharp structure in $N(\varepsilon)$ near the Fermi energy in the electronic band structure [59]. However, in contrast the temperature dependences of $\chi$ and K in $Nb_3Al$ are essentially absent [10]. At least some [34] band structure calculations (see section 2) result in sharp structure in $N(\varepsilon)$ at the Fermi energy also for $Nb_3Al$, which argues for caution in explaining the magnetic data. However, another calculation [59], published at the same time, states that there is no sharp structure in $N(\varepsilon)$ in $Nb_3Al$ near $\varepsilon_F$ which is then consistent with the observed lack of temperature dependence in $\chi$ and K. Possibly the explanation for this disagreement is the degree and positioning with respect to the Fermi energy of the calculated sharp structure in $N(\varepsilon)$.

Unlike unconventional superconductors like the cuprates or the iron based superconductors, there is in general no ordered magnetic behavior in the A15 compound superconductor phase diagrams.

<u>C</u>: The specific heat, C, of the A15's has been thoroughly studied. In the normal state, $C/T = \gamma + \beta T^2 + \delta T^4$; such data give information on the lattice stiffness (the Debye temperature $\Theta_D \propto \beta^{-1/3}$), the electronic density of states at the Fermi energy ($N(0)(1+\lambda) \propto \gamma$ where $\lambda$ is the electron-phonon coupling constant), and the relative strength of the electron-phonon coupling (via the discontinuity in the specific heat at $T_c$, $\Delta C$, divided by $\gamma T_c$). Several reviews contain a section on specific heat data for the A15's, e. g. [44], [54], and [60] and a number of more specialized papers focus on specific heat results for the A15's, e. g. [61]-[64]. For representative values see Table 1.

Table 1: Parameters (except where noted from ref. 61) for Selected A15's, $T_c > 17$ K. Density of states values, rather than using the units states/eV-spin-unit cell used in band structure calculations, are stated in states/eV-atom – more commonly used when discussing specific heat data. There are eight atoms/unit cell and two spins/atom, so 1.8 states/eVatom=7.2 states/eV-spin-unit cell.

| | $T_c$ (K) | $H_{c2}$ (T) | $\gamma$ (mJ/molK$^2$) | $\Theta_D$ (T→0/T>$T_c$) (K) | $\lambda$ | $\Delta C/\gamma T_c$ | $2\Delta/kT_c$ | N(0) exper./theory states/eVatom |
|---|---|---|---|---|---|---|---|---|
| Nb$_3$Ge | 21.8 | 38 [84] | 30.3±1, 34±1 [71] | 302±3 | 1.7±0.2 | 1.9 [71] | 4.2 [65] | 1.2±0.1, 1.5±0.1 [71]/ 1.9[34], 1.8[35] |
| Nb$_3$Ga | 19.8 | 35 [84] | 46±8 | 280/262 | 1.7±0.2 | | | 1.8±0.4/1.8 |
| Nb$_3$Al | 18.7 | 35 [84] | 36±2 | 283±5 | 1.7±0.2 | 2.1 [66] | | 1.4±0.2/1.8[34,35] |
| Nb$_3$Sn* | 17.9 | 21.5 at 1.5 K [70] | 35±3 | 208/270 | 1.7±0.2; 1.8 [68], 1.6±0.1 [69] | | 4.2-4.4 [67-68] | 1.4±0.2/1.5 |
| Nb$_3$Si | 18.0 | | 24±6 | 310±40 | 1.7±0.2 | | | 0.95±0.3/0.6 |
| V$_3$Si* | 16.8* | 19*,22 [70] | 53* | 291-324*/335[70] | 1.29±0.2; 0.96 [68] | 2.0* | 3.5±0.2 [67-68] | 2.4±~0.3/1.8 |
| Nb$_3$Al$_{0.8}$Ge$_{0.2}$ | 20.0 | 43 [84] | 35±2 | 278±5 | 1.7±0.2 | | | 1.4±0.2 |

*transforming

Let us focus on the specific heat of one A15 compound, V$_3$Si, as an example. As seen in Fig. 4, the simple Debye law model for the specific heat, where C$_{lattice}$ is given by only a term cubic in temperature, does not hold above T$_c$ in transforming V$_3$Si.

(Such a simple Debye law does in fact hold above $T_c$ in many A15's, including non-transforming [70] $V_3Si$.) Thus, a plot of C/T (=$\gamma + \beta T^2$ in the simple Debye model) vs $T^2$ in the normal state in Fig. 4 is not a straight line and the Debye temperature must be determined by low enough temperature data that the anharmonic lattice terms in the specific heat (C/T $\propto T^4$ and higher order terms) are negligible. Since data in magnetic fields high enough to suppress superconductivity and to extend the normal state to such low temperatures are rare, in general $\Theta_D$ is determined in the superconducting state at low enough temperature that the superconducting electronic contribution ($\propto \exp(-\Delta/kT)$ where $\Delta$ is the superconducting energy gap, see Table 1) is negligible. See Table 1 for representative values for the Debye temperature in the higher $T_c$ A15's.

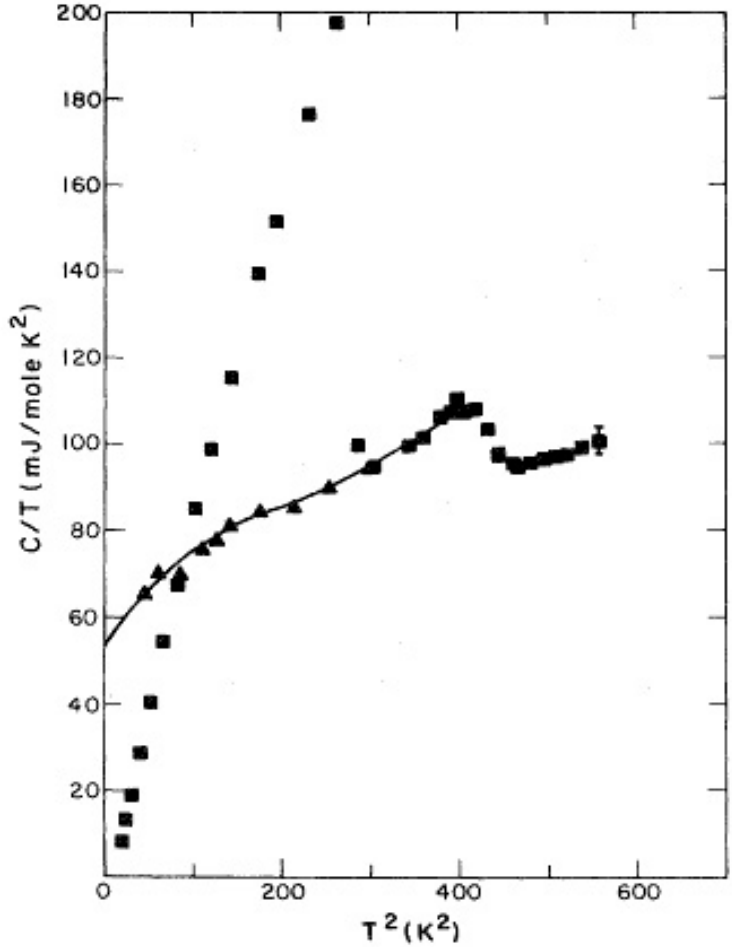

**Fig. 4** Low temperature specific heat divided by temperature, C/T, vs $T^2$ for a single crystal of transforming (martensitic anomaly at 21.2 K, $T_c$=16.8 K) $V_3Si$ [70]. The squares are data in 0 field, the triangles are data measured in 18 T in order to suppress the superconducting transition and better ascertain the extrapolation of $C_{normal}/T$ (T→0), equal to $\gamma$. The large discontinuity in C at $T_c$, $\Delta C$, divided by $\gamma T_c$ is listed in Table 1 as 2.0, indicating strong coupling.

In addition to measuring the specific heat of A15's in high enough magnetic fields to suppress $T_c$ substantially and thus better determine the normal state $\gamma$ (as shown in Fig. 4 for transforming $V_3Si$ as an example), specific heat measurements at lower magnetic fields in the superconducting mixed state have also been carried out. Measurements [70] of the specific heat of transforming $Nb_3Sn$ and $V_3Si$ and non-transforming $V_3Si$ in fields (4, 7, 11, 12.5 T) below $H_{c2}$ were interpreted as showing the straightforward linear increase in C/T (T→0) in the mixed state with field that extrapolated to the normal state γ at $H_{c2}$. A decade later after the advent of the cuprate high temperature superconductors, more careful measurements [72] in non-transforming $V_3Si$ showed a downwards curvature in C/T (T→0) with increasing field below 2 T. This was interpreted not as an indication of nodal behavior (which was the interpretation [73] of such sub-linear increase of C/T (T→0) with field in YBCO) but rather as due to flux line interactions. This 'non-exotic' interpretation was strengthened by further specific heat in field work on the two gap superconductor $NbSe_2$ [74] as well as muon spin resonance measurements in $V_3Si$ [75].

In order to utilize the normal state electronic contribution to C/T, i. e. the γ value, to determine the electronic density of states at the Fermi energy, N(0), (and to compare the experimental value with the theoretical one discussed in section 2), one needs the electron-phonon coupling constant λ (see, e. g., the discussion in [61]).

$$\lambda = 2\int \alpha^2(\omega)F(\omega)\omega^{-1}d\omega$$ where F(ω) is the phonon density of states and α(ω) is the frequency dependent electron-phonon interaction

A common method (see [76] for a discussion in PbBi alloys and [77] for the theory) of experimentally determining the electron-phonon spectral function $\alpha^2(\omega)F(\omega)$ and thus λ is via electron tunneling measurements. These can be supplemented by inelastic neutron scattering measurements of the phonon spectrum F(ω) (see [48] for work on $Nb_3Sb$). Tunneling measurements in the A15's exist (for $Nb_3Sn$, see [67],

[69], [78-79]; for Nb$_3$Ge, see [71]), but can, particularly in the case of thin film samples, be affected by the sample quality at the surface. This, combined with the difficulty of the measurements and thus the incomplete results, resulted in the use of theoretical formulas [32,38] in the study of the A15's to calculate $\lambda$, starting with an inversion of the McMillan formula [32] (discussed above in Section II. 2) put forward in 1968:

$$\lambda = \{1.04 + \mu^* \ln[\Theta_D/1.45T_c]\}/\{(1-0.62\mu^*) \ln[\Theta_D/1.45T_c] - 1.04\}$$

where $\mu^*$ is taken as 0.13. The McMillan formula is known [80-81] to have increasing errors for larger values of $\lambda$, particularly for very strong coupling $\lambda$>1.5. Let us consider two examples using data from Table 1. The McMillan formula for transforming V$_3$Si from Fig. 4 (from Table 1, $\Theta_D$=324 K and $T_c$=16.8 K) gives $\lambda$ = 1.03, while the detailed theoretical calculation by Klein, Boyer, and Papaconstantopoulos [59] gives $\lambda$ = 1.18. Far infrared studies [82], another method for experimentally determining $\lambda$, give $\lambda$ = 1.29. As a second example, consider Nb$_3$Sn from Table 1 ($\Theta_D$=270 K and $T_c$=17.9 K): the inverted McMillan equation gives $\lambda$=1.21 and [59] gives $\lambda$ = 1.12 vs experimental tunneling results (see Table 1) of $\lambda$=1.8 [68] and 1.6 [69]. These examples confirm that the McMillan formula must be used with caution for large $\lambda$.

As discussed above in section 2, $\lambda$ is not the sole parameter important for determining $T_c$. As a comparison to some elemental superconductors, $\lambda$=1.55 for Pb ($T_c$=7.2 K) and $\lambda$=2.13 for Pb$_{0.65}$Bi$_{0.35}$ ($T_c$=8.95 K). [81]

The strong coupling nature of the A15's is further borne out by the normalized energy gap parameters, $2\Delta/kT_c$, listed in Table 1 (where weak coupled BCS theory predicts 3.52) and by the normalized jump in the specific heat, $\Delta C/\gamma T_c$, predicted to be 1.43 in weak coupled BCS theory. (It is interesting to note that the strong coupled, $\Delta C/\gamma T_c = 2.0$, value for transforming $V_3Si$ contradicts the weaker coupled $\lambda$ value and the weak coupling $2\Delta/kT_c=3.5$ value.)

Using the values for $\lambda$ listed in Table 1, and the measured specific heat $\gamma$ values allows the calculation of $N(0)$. Table 1 shows reasonably good agreement between the values determined from $\gamma$ and calculated from band structure calculations. To put these values in context, $N(0)$ for Nb ($T_c=9.2$ K) is [36] 1.4 states/eV-atom. Thus, as discussed in section 2 above, the belief in the late 1960's that a large $N(0)$ due to the 1 dimensional transition metal chains with rather close interatomic spacing was responsible for the high Tc's in some A15's was not born out. This was also made clear by Bednorz and Muller's succeeding discovery of higher $T_c$ in significantly lower $N(0)$ cuprate materials. As a further example, $MgB_2$, $T_c=40$ K, has [83] an $N(0)$ of about 0.8 states/eV-atom.

$\underline{H_{c2}}$: The upper critical magnetic field as a function of temperature, $H_{c2}(T)$ for some representative A15's, as well as for the alloy NbTi, is shown in Fig. 5. [84] As discussed above in section 1, these high upper critical magnetic fields in the A15's, combined with the ability to make practical conductors, are extremely useful in winding solenoids to provide high magnetic fields.

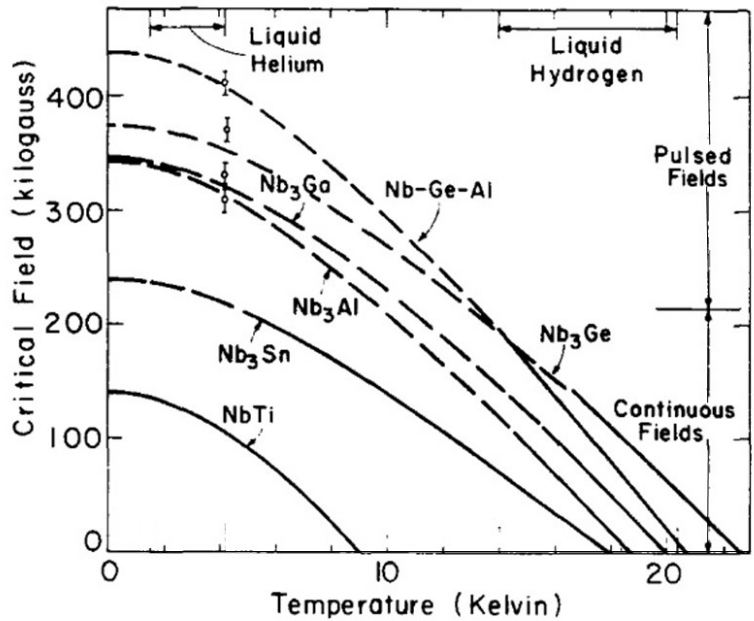

**Fig. 5:** Upper critical magnetic field, $H_{c2}$, as a function of temperature for selected A15 compounds and the alloy NbTi. [84] Measurements in pulsed fields above 21 T are indicated by individual points with error bars for the four highest critical field compounds. The dashed theoretical extrapolations assume no Pauli paramagnetic limiting which agrees fairly well with the pulsed field data.

**Other properties:** One of the more interesting results on $V_3Si$ was the NMR measurement [85] of $1/T_1 \propto T^3$ (usually taken as an indication of nodes, or zeroes, in the superconducting gap). This – combined with the sub-linear behavior [72] of C/T (T→0) with increasing field below 2 T discussed above – led at least for some period of time to the urban legend that $V_3Si$ was an unconventional superconductor similar to the cuprates. However, both the original paper [72] on the sub-linear field response of the specific heat and the NMR paper [85] made clear that there was a preferred, non-exotic superconductor interpretation. In the case of the $1/T_1 \propto T^3$ data, ref. [85] pointed out that this temperature dependence of the spin lattice relaxation rate in $V_3Si$ extended down only to $T_c/3$ and could be fit with a narrow band at the Fermi energy model, as is (see discussion of band structure calculations above) consistent with A15 materials.

The response of $T_c$ to (approximately) hydrostatic pressure in single crystal $V_3Si$ up to 24/18 kbar is [86]/[87] an increase of $+3.7 \times 10^{-5}$ K/bar (certainly smaller than the record rate [88] of $dT_c/dP$ in the first discovered cuprate superconductor, La-Ba-Cu-O, of $10^{-3}$ K/bar). The bulk modulus for $V_3Si$ is [89] 1760 kbar. The martensitic transformation temperature in $V_3Si$, $T_m$, decreases [87] with hydrostatic pressure at a rate of $-1.5\pm0.1 \times 10^{-4}$ K/bar. For $V_3Ga$ ($T_c$=13.9 K) and $V_3Ge$ ($T_c$=6.1 K) the results for $T_c$ vs hydrostatic pressure were [86] $+1.0 \times 10^{-5}$ K/bar and $+8.1 \times 10^{-5}$ K/bar respectively. The thermal electric power of $V_3X$ (X=Ge, Si, Sn, and Ga) was measured from 4 to 300 K and was found to be positive [90]. The thermal conductivity of $Nb_3Sn$ was measured between 2 and 86 K and showed very little temperature dependence above $T_c$ [91]. A Hall effect study of $Nb_3Sn$ up to 300 K, using a one band model, found it to be p-type, with approximately ¼ carrier/atom [92]. de Haas van Alphen studies of $Nb_3Sn$ above $T_c$ [92] and $V_3Si$ both above [93] and below [94] $T_c$ show only a limited number of orbits due to both relatively high residual resistivities (RRR for the $Nb_3Sn$ was ≈ 50) and smearing effects due to the martensitic transformation present in both. However the orbits seen agree [93] with band structure calculations. Positron annihilation experiments, together with band structure calculations, in $Nb_3Sn$ have been used [95] to map out the Fermi surfaces, with the result that a structure in one of the six bands that cross the Fermi surface shows a very high density of states from the Nb 4 d electrons. The authors theorize that these high density of states electrons are important for the 18 K superconductivity. The superconducting coherences lengths for $Nb_3Sn$ and $V_3Si$ are [68] both about 55-60 Å (at least a factor of three longer than in the cuprates), while the zero temperature London penetration depth for $Nb_3Sn$ is approximately 1000 Å. [68]

**4. Attempts to exceed $T_c$=23 K in $Nb_3Ge$: A15 $Nb_3Si$:**

Matthias and co-workers [22] in the mid-1960's, in $Nb_3Ge$, were already following the idea that the maximum $T_c$ in a *given* compound occurs at the ideal 3:1 stoichiometry. (A good example of this, as discussed above, is $V_3Ga$, see refs. [21] and [96]). Useful in this effort was the experimentally based (on work with $Nb_3Al_{0.8}Ge_{0.2}$) prediction [97] what the stoichiometric $Nb_3Ge$ lattice parameter should be: 5.154 Å. (Note the theoretical prediction of Geller [27], $a_0$=5.12 Å.) After Gavaler [13-14] succeeded in stabilizing A15 $Nb_3Ge$ via sputtering with a lattice parameter of 5.152 ± 0.001 Å and $T_c \approx$ 23 K, Geller - using phenomenological arguments – predicted [98] in 1975 that $T_c$ = 31-35 K for $Nb_3Si$, $a_0$=5.06 Å. Noolandi and Testardi [99] in 1977 – based on a correlation of decreasing $T_c$ with increasing lattice parameter (caused by two effects which they analyzed to be equivalent: off-stoichiometry or defects due to neutron irradiation) in a large number of A15's – predicted $T_c \geq$ 25 K for <u>defect-free</u> A15 $Nb_3Si$, with $a_0$ predicted to be 5.08 Å. The hunt for higher $T_c$ in the A15's then switched from the decades long effort to maximize $T_c$/achieve 3:1 stoichiometry/minimize defects in $Nb_3Ge$ to trying to prepare $Nb_3Si$ (which forms out of the melt in the tetragonal $Ti_3P$ type structure) in the denser (unit cell volume is smaller by 2.8%) A15 structure. Efforts focused on high pressure treatments to stabilize $Nb_3Si$ in the more compact A15 atomic arrangement.

Initial efforts to prepare A15 $Nb_3Si$ by explosive compression at 1 Mbar, by Pan et al. [100] in 1975 and Dew Hughes and Linse [101] in 1979, were made. Although resistive onsets of superconductivity were obtained around 19 K, no convincing x-ray evidence of bulk production of the A15 structure in $Nb_3Si$ was obtained. In 1981, Olinger and Newkirk [102] succeeded, using explosive compression at 1 Mbar, in making a sample consisting of 50-70% A15 $Nb_3Si$, $a_0$=5.091 ± 0.006 Å, with an inductive transition at 18.5 K and a bulk discontinuity ($\Delta C/\gamma T_c$=2.0) starting at 18.0 K in the specific heat [103]. Presumably the unavoidable presence of defects caused by the extremely non-equilibrium preparation method contributed to a suppression of the $T_c$ from the higher $T_c$ predictions. Although there were further attempts to improve on $T_c$ in A15 $Nb_3Si$ prepared under high pressure (see e. g. [104]), no significant increase in $T_c$ was achieved. For higher superconducting transition temperatures, attention shifted to the high $T_c$ cuprates in 1986.

One recent development, the discovery in 2008 of $T_c$=38 K in A15 $Cs_3C_{60}$ [105] under 7 kbar pressure, serves as a contrast to what has been described above in this review. $Cs_3C_{60}$, where the Cs atoms occupy the 1 dimensional chain sites on the cube faces (fig. 1) and the fulleride $C_{60}$ occupies the body centered site and the cube corners, has an enormous lattice parameter (11.78 Å) so that the A-atom (in this case Cs) interatomic spacing, $a_0/2$=5.89 Å, along the chains is over 10% *larger* than the interatomic spacing (5.32 Å) in the pure bcc Cs metal. Why $Cs_3C_{60}$ forms in the A15 structure, while the other alkali $A_3C_{60}$ compounds form in the fcc structure, is presumably linked to the significantly larger size of Cs vs Rb (4.94 Å) or K (4.61 Å) and the accompanying decrease in overlap of the fulleride anions. This larger size also leads to $Cs_3C_{60}$ being an insulator at ambient pressure, while the other, fcc alkali $A_3C_{60}$ compounds are conducting.

## 5. Comparison of A15's with other superconductor classes; Summary

Comparisons of the properties of the A15's to other superconductors have been made throughout this review. A15's, despite the results in $V_3Si$ (NMR $1/T_1 \propto T^3$ behavior [85] down to $T_c/3$ and sub-linear increase [72] of $C/T$ ($T \rightarrow 0$) in the superconducting state with field) reminiscent of nodal behavior in the cuprate and iron based superconductors, are commonly believed to be strongly coupled electron-phonon coupled s-wave superconductors. The coupling strength, as evinced by $2\Delta/kT_c$ and $\Delta C/\gamma T_c$ values (see Table 1) enhanced over the BCS values of 3.52 and 1.43 respectively, of the superconducting pairing in the higher $T_c$ A15's is certainly strong compared to the elemental superconductors but similar, despite their lower (0.5 – 2.3 K) $T_c$ values, to the heavy Fermion superconductors. The structure of the A15's is cubic and 3-dimensional, with very little anisotropy in the properties, strongly contrasting with the layered cuprates.

Concerning applications of superconductivity, cuprate superconductors are used for cell phone tower noise filters and are beginning [106] to be produced in wire form for superconducting high field (>30 T) magnets. However, the workhorse of superconducting magnets for fields above 10 T (including the ITER fusion project and high field NMR magnets) continues to be doped A15 $Nb_3Sn$, with over $10^7$ m of wire produced annually. Thus, the discovery [7] in 1954 of superconductivity in A15 $Nb_3Sn$ by Matthias, Geballe, Geller and Corenzwit continues to be of importance to modern society today.

Acknowledgements: Work at Florida supported by the U S Department of Energy, Office of Basic Energy Sciences, Division of Materials Sciences and Engineering, contract no. DE-FG02-86ER45268. Helpful discussions and input for this review from M. L. Cohen, W.

**Pickett, and B. Klein are gratefully acknowledged. While the author was a guest at the Kavli Institute for Theoretical Physics in Santa Barbara, comments on A15's by F. Ronning, P. Coleman, and S. Kivelson were helpful. Fruitful collaboration on A15's with A. L. Giorgi, L. R. Newkirk, G. W. Webb, A. R. Sweedler, C. W. Chu and B. T. Matthias between 1978 and 1983 formed the basis of this review. For Figures 3/4, copyright 1966/1984 by the American Physical Society.**